\documentclass[12pt]{article}
\usepackage{graphicx}
\usepackage{amsmath}

\newtheorem{theorem}{Theorem}

\input{tcilatex}

\begin{document}

\title{Integrable Systems and Metrics of Constant Curvature}
\author{Maxim Pavlov \\
Loughborough University, UK}
\maketitle


\textit{keywords}: Lagrangian, metrics of constant curvature, Hamiltonian
structure, reciprocal transformation, Poisson brackets.

\textbf{Abstract.}

In this article we present a Lagrangian representation for evolutionary
systems with a Hamiltonian structure determined by a differential-geometric
Poisson bracket of the first order associated with metrics of constant
curvature. Kaup-Boussinesq system has three local Hamiltonian structures and
one nonlocal Hamiltonian structure associated with metric of constant
curvature. Darboux theorem (reducing Hamiltonian structures to canonical
form ''d/dx'' by differential substitutions and reciprocal transformations)
for these Hamiltonian structures is proved.

\section{Introduction}

In this article we describe nonlocal Hamiltonian structure associated with
differential-geometric Poisson bracket of the first order with metric of
constant curvature and its Lagrangian representation for evolutionary systems

\begin{equation}
u_{t}^{k}=f^{k}(\mathbf{u,u}_{x},...).  \tag{1.1}
\end{equation}
It means that (1.1) can be re-written as
\begin{equation}
u_{t}^{i}=\{u^{i},H\}=\overset{\wedge }{A}^{ik}\frac{\delta H}{\delta u^{k}},
\tag{1.2}
\end{equation}
where $\overset{\wedge }{A}^{ik}$is a Hamiltonian operator, $H=\int h(%
\mathbf{u,u}_{x},...)dx$ is a functional of conservation law density $h(%
\mathbf{u,u}_{x},...)$ and $\{u^{i}(x),u^{j}(x^{\prime })\}$ is a Poisson
bracket. Then we can introduce new variables $a^{\alpha }(\mathbf{u}%
)=\partial _{x}\varphi ^{\alpha }$, where the system (1.1) is determined by
action
\begin{equation}
S=\int L(\mathbf{\varphi }_{t}\mathbf{,\varphi }_{x}\mathbf{,\varphi }_{xx}%
\mathbf{,\varphi }_{xxx},...)dxdt,  \tag{1.3}
\end{equation}
where $L(\mathbf{\varphi }_{t}\mathbf{,\varphi }_{x}\mathbf{,\varphi }_{xx}%
\mathbf{,\varphi }_{xxx},...)$ is a Lagrangian.

The modern theory of Hamiltonian and Symplectic structures, Poisson brackets
and Lagrangian representations is developed in [1] by L.D.Faddeev and
V.E.Zakharov in 1971, where they showed that the Korteweg-de Vries equation
has Poisson bracket
\begin{equation}
\{a(x),a(x^{\prime })\}=\partial _{x}\delta (x-x^{\prime })  \tag{1.4}
\end{equation}
determining the first Hamiltonian structure
\begin{equation}
a_{t}=\partial _{x}\frac{\delta H}{\delta a},  \tag{1.5}
\end{equation}
where (in general case) the Hamiltonian is $H=\int h(\mathbf{a,a}_{x},...)dx$%
. $N$-component generalization of this formula on arbitrary dependence $u=u(%
\mathbf{a}(x))$ was established in the article [2] by B.A.Dubrovin and
S.P.Novikov in 1983

\begin{equation}
\{u^{i}(x),u^{j}(x^{\prime })\}=[g^{ij}(\mathbf{u}(x))\partial
_{x}-g^{is}\Gamma _{sk}^{j}u_{x}^{k}]\delta (x-x^{\prime }),  \tag{1.6}
\end{equation}
where $g^{ij}(\mathbf{u})$ is nondegenerated symmetric flat metric, $\Gamma
_{jk}^{i}$ are the coefficients of the corresponding Levi-Civita connection,
$\Gamma _{sk}^{j}=\Gamma _{ks}^{j}$ and $\nabla _{i}g_{sk}=0$. If we choose
the Hamiltonian depended on functions $u^{i}$ only, $H=\int h(\mathbf{u})dx$%
, then the Poisson bracket (1.6) determines Hydrodynamic type system $%
u_{t}^{i}=w_{k}^{i}(\mathbf{u})u_{x}^{k}$, where $w_{k}^{i}(\mathbf{u}%
)=\nabla ^{i}\nabla _{k}h$ (see [3]). Moreover, we can find ''flat
coordinates'' $a^{\nu }$ (annihilators of the Poisson bracket (1.6), or
Casimirs), where all $\Gamma _{\beta \gamma }^{\alpha }\equiv 0$ and $%
\overset{\_}{g}^{\alpha \beta }$is constant symmetric nondegenerated metric
\begin{equation}
a_{t}^{\alpha }=\partial _{x}[\overset{\_}{g}^{\alpha \beta }\frac{\delta H}{%
\delta a^{\beta }}].  \tag{1.7}
\end{equation}
This Hamiltonian structure allows ($N+2$) conservation laws, where first $N$
of them are (1.7), the conservation law of Energy is
\begin{equation}
h_{t}=\partial _{x}[\overset{\_}{g}^{\alpha \beta }\frac{\partial h}{%
\partial a^{\alpha }}\frac{\partial h}{\partial a^{\beta }}],  \tag{1.8}
\end{equation}
and conservation law of Momentum $P=\int pdx$ is
\begin{equation}
p_{t}=\partial _{x}[a^{\alpha }\frac{\partial h}{\partial a^{\alpha }}-h],
\tag{1.9}
\end{equation}
where ($\overset{\_}{g}_{\alpha \beta }\overset{\_}{g}^{\beta \gamma
}=\delta _{\alpha }^{\gamma }$)

\begin{equation}
p=\frac{1}{2}\overset{\_}{g}_{\alpha \beta }a^{\alpha }a^{\beta }.
\tag{1.10}
\end{equation}
In more general case $H=\int h(\mathbf{a,a}_{x}\mathbf{,...,a}_{M})dx$ the
conservation law of the Momentum is
\begin{equation}
p_{t}=\partial _{x}[a^{\alpha }\frac{\delta H}{\delta a^{\alpha }}-F],
\tag{1.11}
\end{equation}
where
\begin{equation}
\partial _{x}F=\frac{\delta H}{\delta a^{\alpha }}a_{x}^{\alpha }.
\tag{1.12}
\end{equation}
and thus
\begin{equation*}
F=h-\overset{M}{\underset{n=1}{\sum }}(-1)^{n}a_{n}^{\beta }\overset{M}{%
\underset{k=n}{\sum }}(-1)^{k}\partial _{x}^{k-n}\frac{\partial h}{\partial
a_{k}^{\beta }}.
\end{equation*}
In this case formula (1.8) are transformed into more general
\begin{equation*}
h_{t}=\partial _{x}(\overset{\_}{g}^{\alpha \beta }[\overset{M}{\underset{m=1%
}{\sum }}\overset{m-1}{\underset{k=0}{\sum }}(-1)^{k}\partial _{x}^{k}(\frac{%
\partial h}{\partial a_{m}^{\alpha }})\partial _{x}^{m-k}(\frac{\delta H}{%
\delta a^{\beta }})+\frac{1}{2}\frac{\delta H}{\delta a^{\alpha }}\frac{%
\delta H}{\delta a^{\beta }}]).
\end{equation*}
It is easily to check that if evolutionary system (1.1) has ($N+1$)
conservation law densities connected by constraint (1.10), then this system
has local Hamiltonian structure (1.7). Also, the evolutionary system (1.7)
has the Lagrangian representation

\begin{equation}
S=\int [\frac{1}{2}\overset{\_}{g}_{\alpha \beta }\varphi _{x}^{\alpha
}\varphi _{t}^{\beta }-h(\mathbf{\varphi }_{x}\mathbf{,\varphi }%
_{xx},...)]dxdt\text{,}  \tag{1.13}
\end{equation}
where $a^{\alpha }=\varphi _{x}^{\alpha }$.

A more complicated case was studied by E.V.Ferapontov and O.I.Mokhov in the
article [4] in 1990:
\begin{equation}
\{u^{i}(z),u^{j}(z^{\prime })\}=[g^{ij}(\mathbf{u}(z))\partial
_{z}-g^{is}\Gamma _{sk}^{j}u_{z}^{k}+\varepsilon u_{z}^{i}\partial
_{z}^{-1}u_{z}^{j}]\delta (z-z^{\prime }),  \tag{1.14}
\end{equation}
where $g^{ij}(\mathbf{u})$ is nondegenerated symmetric metric with constant
curvature $\varepsilon $ (see (1.6)). However, some problems have been
unsolved. In this article we present:

\begin{enumerate}
\item Canonical coordinates for evolutionary systems with nonlocal
Hamiltonian structure \qquad determined by the Poisson bracket (1.6). Thus,
Hamiltonian structure will be written in compact form (see for comparison
(1.7)).

\item The Metric and the Momentum in canonical coordinates (see for
comparison (1.9) and (1.10)).

\item The Lagrangian representation (see for comparison (1.13)).

\item Reciprocal transformations connecting Poisson brackets (1.6) and
(1.14).

\item The fourth (nonlocal) Hamiltonian structure associated with metric of
constant curvature for the Kaup-Boussinesq system.
\end{enumerate}

Local linear-degenerated Lagrangians (Lagrangians are linear with respect to
derivatives of $t$) were studied in [8]. It means that symplectic structure
is local (determined by differential operator of arbitrary order), of
course, corresponding Hamiltonian structure is nonlocal, but not invertible
in compact form (it means that in general case corresponding differential
operator has infinite set of elements). Arbitrary nonlocal Hamiltonian
structure has corresponding nonlocal symplectic structure. Moreover this
symplectic structure has infinite set of elements too. It is astonish that
namely in case of constant curvature metric nonlocal Hamiltonian structure
has local corresponding symplectic structure. By another words, not every
nonlocal Hamiltonian structure has inverse local symplectic structure. At
first the general reciprocal transformation connecting Poisson brackets
(1.6) and (1.14) was presented by E.V.Ferapontov (see below). However, here
we present one very special case: one-parametric family of constant
curvature metrics $\varepsilon -$intimated to flat case. It means that if in
below presented reciprocal transformation anyone put $\varepsilon =0$, then
it will be identical transformation (Moreover in general case recalculation
of all attributes for Poisson bracket associated with metric of constant
curvature (annihilators, momentum and Hamiltonian) is very complicated
problem not solved now. Just in our particular case it was solved and
presented below). The metric of constant curvature is well known, however
annihilators and momentum for corresponding Poisson brackets were not known
as well as Lagrangian representations. Our major aim is construction of
Lagrangian representations without constraints for nonlocal Hamiltonian
structures. Here we establish a Lagrangian representation for nonlocal
Hamiltonian structure associated with differential-geometric Poisson bracket
of the first order with metric of constant curvature. This is the first
nontrivial example of nonlocal Hamiltonian structures generalizing the local
one. Lagrangian representations for Poisson brackets associated with metrics
of constant curvature have been obtained by application of special
reciprocal transformation (see below) for Lagrangian representations of
Poisson brackets associated with metrics of zero curvature. It is amazing,
that it is possible. Usually it is not valid. If anyone try to apply
arbitrary reciprocal transformation for arbitrary Lagrangian density (which
is 2-form), then obtained new Lagrangian representation will not create
system connected by abovementioned reciprocal transformation with initial
system determined by initial Lagrangian representation. It means that we do
not know all Lagrangian representations convertible under reciprocal
transformation into others. However, namely in case of constant curvature
metric this problem is successfully solved in this article. Moreover, we
would like to emphasize that in case of constant curvature knowledge of
annihilators and momentum is not enough for direct reconstruction of
Lagrangian representation with respect to Hamiltonian structures associated
with metrics of zero curvature. This is nontrivial problem are solved by
specific choice of annihilators (special $N$ from all $N+1$, see below).
This article contains several Sections. In Section II we formulate a theorem
about canonical coordinates (Casimirs or annihilators of Poisson brackets),
where this nonlocal Hamiltonian structure will be compactly presented. In
Section III we present two theorems about relationship between this nonlocal
and local Hamiltonian structures. In the Section IV we present two
remarkable examples, which allow this nonlocal Hamiltonian structure. One of
them is the Calogero KdV equation related to the KdV equation by the
combination of differential substitutions, another is the Thrice-Modified
Kaup-Boussinesq system which is related to the Kaup-Boussinesq system by a
combination of differential substitutions. In Section V we establish a
Lagrangian representation for an arbitrary evolutionary system with this
nonlocal Hamiltonian structure. And we show that canonical coordinates
presented in Section II determine potential functions in this Lagrangian
representation. Moreover, we show relationship between this Lagrangian
representation (for evolutionary system with nonlocal Hamiltonian structure)
and with Lagrangian representation for evolutionary system determined by
local Hamiltonian structure. In Section VI we describe a very important
example of the first four Hamiltonian structures of the Kaup-Boussinesq
system. We demonstrate validity of infinite-dimensional analog of Darboux
theorem for this Hamiltonian structures by straightforward calculations,
where every Hamiltonian structure can be presented in their canonical form
''d/dx''. In all cases we present Lagrangian representations, describe
relationships between all formulas, and present a new integrable
evolutionary system connected with Thrice-Modified Kaup-Boussinesq system by
reciprocal transformation, which has local Hamiltonian structure reduced
from nonlocal Hamiltonian structure of aforementioned type.

To this moment we know many integrable systems possessing this nonlocal
Hamiltonian structure. We mention here just some famous of them. These are
Korteweg-de Vries equation, Kaup-Boussinesq system, Multi-component
Long-Short Wave Resonance (see articles of Najima \& Oikawa and Melnikov),
Coupled KdV (see articles of Antonowicz \& Fordy) and so on. Moreover,
averaged integrable systems are hydrodynamic type systems (see articles of
Dubrovin \& Novikov), which possess the same type of Hamiltonian structures.
The modern level of development of Hamiltonian structures (see below) needs
for introducing them into other areas of scientific creation like fields
theory, theory of instability in fluid mechanics and ets. We hope that
presented results can be interesting for specialists not working in theory
of integrable systems or in differential geometry as well.

\section{Canonical Coordinates for the Metrics of Constant Curvature}

\begin{theorem}
The evolutionary system (1.1) (see (1.14))
\begin{equation}
u_{y}^{i}=[g^{ij}\partial _{z}-g^{is}\Gamma _{sk}^{j}u_{z}^{k}+\varepsilon
u_{z}^{i}\partial _{z}^{-1}u_{z}^{j}]\frac{\delta H}{\delta u^{j}}\text{, \
\ }i=1,2...N  \tag{2.1}
\end{equation}

has

\begin{enumerate}
\item Casimir functionals $H_{\alpha }=\int c^{\alpha }(\mathbf{u})dz$,
where $\alpha =1,2...N$ (annihilators of the Poisson bracket (1.14)), which
are determined by (see (2.1))
\begin{equation}
\lbrack g^{ij}\partial _{z}-g^{is}\Gamma _{sk}^{j}u_{z}^{k}+\varepsilon
u_{z}^{i}\partial _{z}^{-1}u_{z}^{j}]\frac{\delta H_{\alpha }}{\delta u^{j}}%
=0,  \tag{2.2}
\end{equation}
or by
\begin{equation}
\partial _{ik}c^{\alpha }-\Gamma _{ik}^{n}\partial _{n}c^{\alpha
}+\varepsilon g_{ik}c^{\alpha }=0.  \tag{2.3}
\end{equation}
(the system (2.3) has ($N+1$) solutions. Any $N$ of them are functionally
independent.)

\item The metric $g^{\alpha \beta }$ in Casimirs $c^{\alpha }(u)$ (see
(2.3)) is
\begin{equation}
g^{\alpha \beta }=\overset{\_}{g}^{\alpha \beta }-\varepsilon c^{\alpha
}c^{\beta },  \tag{2.4}
\end{equation}
where $\overset{\_}{g}^{\alpha \beta }$is nondegenerated symmetric constant
matrix. The metric $g_{\alpha \beta }$ (see (2.4), where $g_{\alpha \beta
}g^{\beta \gamma }=\delta _{\alpha }^{\gamma }$) is
\begin{equation}
g_{\alpha \beta }=\overset{\_}{g}_{\alpha \beta }+\varepsilon \frac{\overset{%
\_}{g}_{\alpha \gamma }c^{\gamma }\overset{\_}{g}_{\beta \nu }c^{\nu }}{%
1-\varepsilon \overset{\_}{g}_{\gamma \nu }c^{\gamma }c^{\nu }},  \tag{2.5}
\end{equation}
where $\overset{\_}{g}_{\alpha \beta }\overset{\_}{g}^{\beta \gamma }=\delta
_{\alpha }^{\gamma }$. The Christoffel symbols are $\Gamma _{\beta \gamma
}^{\alpha }=\varepsilon g_{\beta \gamma }c^{\alpha }$.

\item The first $N$ conservation laws (see (1.12) and (2.4)) are
\begin{equation}
c_{y}^{\alpha }=\partial _{z}[g^{\alpha \beta }\frac{\delta H}{\delta
c^{\beta }}+\varepsilon c^{\alpha }F]\text{, \ \ \ \ }\alpha =1,2,...N
\tag{2.6}
\end{equation}

\item The conservation law of the momentum
\begin{equation}
p_{y}=\partial _{z}[(1-\varepsilon p)(c^{\alpha }\frac{\delta H}{\delta
c^{\alpha }}-F)]\text{,}  \tag{2.7}
\end{equation}
where the Momentum $P=\int p(\mathbf{c})dx$ is
\begin{equation}
p=\frac{1}{\varepsilon }[1-\sqrt{1-\varepsilon \overset{\_}{g}_{\alpha \beta
}c^{\alpha }c^{\beta }}],  \tag{2.8}
\end{equation}
which is determined by (2.6)
\begin{equation}
g^{\alpha \beta }\partial _{\beta }p+\varepsilon c^{\alpha }p=c^{\alpha }.
\tag{2.9}
\end{equation}

\item The conservation law of the Energy is
\begin{equation}
h_{y}=\partial _{z}(\frac{1}{2}g^{\alpha \beta }\frac{\partial h}{\partial
c^{\alpha }}\frac{\partial h}{\partial c^{\beta }}+\frac{\varepsilon }{2}%
h^{2})  \tag{2.10}
\end{equation}
for hydrodynamic type systems ($H=\int h(\mathbf{c})dz$) or in more general
case (see (2.6))
\begin{equation*}
h_{y}=\partial _{z}(\overset{M}{\underset{m=1}{\sum }}\overset{m-1}{\underset%
{k=0}{\sum }}(-1)^{k}\partial _{z}^{k}(\frac{\partial h}{\partial
c_{m}^{\alpha }})\partial _{z}^{m-k}(g^{\alpha \beta }\frac{\delta H}{\delta
c^{\beta }}+\varepsilon c^{\alpha }F)+\frac{1}{2}g^{\alpha \beta }\frac{%
\delta H}{\delta c^{\alpha }}\frac{\delta H}{\delta c^{\beta }}+\frac{%
\varepsilon }{2}F^{2}),
\end{equation*}
where $H=\int h(\mathbf{c,c}_{z}\mathbf{,...c}_{M})dz$ and $F=h-\overset{M}{%
\underset{n=1}{\sum }}(-1)^{n}c_{n}^{\beta }\overset{M}{\underset{k=n}{\sum }%
}(-1)^{k}\partial _{x}^{k-n}\frac{\partial h}{\partial c_{k}^{\beta }}$. \ \
\ \ \ \ \ \ \ \ \ \ \ \ \ \ \ \ \ \ \ \ \ \ \ \ \ \ \ \ \ \ \ \ \ \ \ \ \ \
\ \ \ \ \ \ \ \ \ \ \ \ \ \ \ \ \ \ \ \ \ \ \ \ \ \ \ \ \ \ \ \ \ \ \ \ \ \
\ \ \ \ \ \ \ \ \ \ \ \ \
\end{enumerate}
\end{theorem}

\textbf{Remark I}: If $H=\int h(\mathbf{c})dz$, then evolutionary system
(2.1) transforms into Hydrodynamic type system $u_{y}^{i}=w_{k}^{i}(\mathbf{u%
})u_{z}^{k}$, where $w_{k}^{i}(\mathbf{u})=\nabla ^{i}\nabla
_{k}h+\varepsilon h\delta _{k}^{i}$ (see [4]).

\textbf{Remark II}: If $\varepsilon \rightarrow 0$, all formulas (2.1-10)
transform into the ''flat'' case (1.6-11).

\textbf{Proof}: can be obtained by straightforward calculation.

\textbf{Example}: Hydrodynamic type systems possessing nonlocal Hamiltonian
structure (2.1) associated with elliptic coordinates were described in [5],
where all exact formulas (Casimirs, metrics, conservation law densities)
were presented too.

\begin{theorem}
If evolutionary system (1.1) has ($N+1$) conservation law densities
connected by constraint (2.8), then this system has nonlocal Hamiltonian
structure associated with metric of constant curvature.
\end{theorem}

\textbf{Proof}: We take evolutionary system (1.1) in divergent form $%
c_{y}^{\alpha }=\partial _{z}b^{\alpha }(\mathbf{c,c}_{z}\mathbf{,c}%
_{zz},...)$, then additional conservation law $p_{y}=\partial _{z}b(\mathbf{c%
})$ yields relationship $\partial _{z}b(\mathbf{c})=\overset{\_}{g}_{\alpha
\beta }\frac{c^{\beta }}{1-\varepsilon p}b_{z}^{\alpha }$. It is valid if
and only if $\overset{\_}{g}_{\alpha \beta }b^{\beta }=\delta S/\delta
q^{\alpha }$, where $q^{\alpha }=c^{\alpha }/(1-\varepsilon p)$ and $S=\int
s(\mathbf{q,q}_{z}\mathbf{,q}_{zz},...)dz$. It means that $p_{y}=\partial
_{z}[q^{\beta }\frac{\delta S}{\delta q^{\beta }}-R]$ and $c_{y}^{\alpha
}=\partial _{z}[\overset{\_}{g}^{\alpha \beta }\frac{\delta S}{\delta
q^{\beta }}]$, where $\partial _{z}R=\frac{\delta S}{\delta q^{\alpha }}%
q_{z}^{\alpha }$. Since $\frac{\delta S}{\delta q^{\alpha }}=(1-\varepsilon
p)[\frac{\delta S}{\delta c^{\alpha }}-\varepsilon c^{\gamma }\frac{\delta S%
}{\delta c^{\gamma }}\overset{\_}{g}_{\alpha \beta }c^{\beta }]$, anyone can
immediately obtain (2.6) and (2.7), where $h=(1-\varepsilon p)s$.

\section{Reciprocal Transformation and nonlocal Hamiltonian structures.}

In this Section we establish canonical reciprocal transformation between
local Hamiltonian structure (1.7) and nonlocal Hamiltonian structure (2.6).

The general reciprocal transformation between Hamiltonian structures (1.7)
and (2.6) was constructed in [6] by E.Ferapontov in 1995 for Hydrodynamic
type systems
\begin{equation}
u_{t}^{i}=\upsilon _{k}^{i}(\mathbf{u})u_{x}^{k}\ \ \ \ \text{and}\ \ \ \ \
u_{y}^{i}=w_{k}^{i}(\mathbf{u})u_{z}^{k}.  \tag{3.1}
\end{equation}

If the first system in (3.1) has local Hamiltonian structure (see (1.7-10)),
then we can introduce the reciprocal transformation
\begin{equation}
dy=A(\mathbf{u})dx+B(\mathbf{u})dt\text{,}\ \ \ \ \ dz=C(\mathbf{u})dx+D(%
\mathbf{u})dt,  \tag{3.2}
\end{equation}
where
\begin{eqnarray*}
A(\mathbf{u}) &=&\alpha h+\beta p+\gamma _{\nu }a^{\nu }+\zeta \text{,}\ \ \
\ B(\mathbf{u})=\frac{\alpha }{2}\overset{\_}{g}^{\mu \nu }h_{\mu }h_{\nu
}+\beta (a^{\nu }h_{\nu }-h)+\gamma _{\nu }\overset{\_}{g}^{\nu \mu }h_{\mu
}+\eta \\
C(\mathbf{u}) &=&\overset{\_}{\alpha }h+\overset{\_}{\beta }p+\overset{\_}{%
\gamma }_{\nu }a^{\nu }+\overset{\_}{\zeta }\text{,}\ \ \ \ D(\mathbf{u})=%
\frac{\overset{\_}{\alpha }}{2}\overset{\_}{g}^{\mu \nu }h_{\mu }h_{\nu }+%
\overset{\_}{\beta }(a^{\nu }h_{\nu }-h)+\overset{\_}{\gamma }_{\nu }\overset%
{\_}{g}^{\nu \mu }h_{\mu }+\overset{\_}{\eta },
\end{eqnarray*}

and $\alpha $, $\beta $, $\gamma _{\nu }$, $\zeta $, $\eta $, $\overset{\_}{%
\alpha }$, $\overset{\_}{\beta }$, $\overset{\_}{\gamma }_{\nu }$, $\overset{%
\_}{\zeta }$, $\overset{\_}{\eta }$ are arbitrary constants.

\begin{theorem}
([6]) The Hydrodynamic type system ($x,t$) with local Hamiltonian structure
(1.7) transforms into the Hydrodynamic type system ($y,z$) with nonlocal
Hamiltonian structure (2.6) if
\begin{equation}
\overset{\_}{g}^{\mu \nu }\gamma _{\mu }\gamma _{\nu }-2\alpha \eta -2\beta
\zeta =\varepsilon ,  \tag{3.3}
\end{equation}
\ \
\begin{equation*}
\overset{\_}{g}^{\mu \nu }\overset{\_}{\gamma }_{\mu }\overset{\_}{\gamma }%
_{\nu }=2\overset{\_}{\alpha }\overset{\_}{\eta }+2\overset{\_}{\beta }%
\overset{\_}{\zeta },\ \ \overset{\_}{g}^{\mu \nu }\gamma _{\mu }\overset{\_}%
{\gamma }_{\nu }=\alpha \overset{\_}{\eta }+\overset{\_}{\alpha }\eta +\beta
\overset{\_}{\zeta }+\overset{\_}{\beta }\zeta
\end{equation*}
\end{theorem}

By choosing special constants in (3.3) we present particular, but more
simple and more clear

\begin{theorem}
The evolutionary system ($x,t$) with local Hamiltonian structure (1.7)
transforms itself into the evolutionary system ($y,z$) with nonlocal
Hamiltonian structure (2.6) by the reciprocal transformation
\begin{equation}
dy=dt,\ \ dz=(1+\frac{\varepsilon }{2}p)dx+\frac{\varepsilon }{2}qdt,
\tag{3.4}
\end{equation}
where $\partial _{t}p=\partial _{x}q$ and $q=a^{\alpha }\frac{\delta H}{%
\delta a^{\alpha }}-F$ (see (1.11) and (1.12)). Then
\begin{eqnarray*}
1\text{. \ \ \ \ \ }\overset{\_}{h}(\mathbf{c,c}_{z},...) &=&h(\mathbf{a,a}%
_{x},...)/(1+\frac{\varepsilon }{2}p), \\
2\text{. \ \ \ \ \ \ \ \ \ \ \ \ \ \ \ \ \ }\overset{\_}{p} &=&p/(1+\frac{%
\varepsilon }{2}p), \\
3\text{. \ \ \ \ \ \ \ \ \ \ \ \ \ \ \ }c^{\alpha } &=&a^{\alpha }/(1+\frac{%
\varepsilon }{2}p), \\
4\text{. \ \ \ \ \ \ \ \ \ \ \ \ \ \ }g^{\alpha \beta } &=&\overset{\_}{g}%
^{\alpha \beta }-\varepsilon c^{\alpha }c^{\beta }, \\
5\text{. \ \ \ \ \ \ \ \ \ \ \ \ \ \ \ }c_{y}^{\alpha } &=&\partial
_{z}[g^{\alpha \beta }\frac{\delta \overset{\_}{H}}{\delta c^{\beta }}%
+\varepsilon c^{\alpha }\overset{\_}{F}],
\end{eqnarray*}
\ \ \ \ where $\partial _{x}=(1+\frac{\varepsilon }{2}p)\partial _{z}$, $%
\overset{\_}{H}=\int \overset{\_}{h}(\mathbf{c,c}_{z},...)dz$ and $\partial
_{z}\overset{\_}{F}=\frac{\delta \overset{\_}{H}}{\delta c^{\beta }}%
c_{z}^{\beta }$
\end{theorem}

\textbf{Remark I}: if $\varepsilon \rightarrow 0$, then all this formulas
transform into local case (see Section I). E.V.Ferapontov have studied
another particular case too
\begin{equation*}
dy=dt\text{, \ \ \ }dz=(p+\frac{1}{2})dx+(a^{\alpha }\frac{\partial h}{%
\partial a^{\alpha }}-h)dt.
\end{equation*}
However, just in our case we describe one-parameter ($\varepsilon -$
parameter) family of metrics of constant curvature, where if $\varepsilon =0$
(3.4) is identical. In our case we present by above theorem recalculation
for annihilators, momentum and Hamiltonian -- all that was absent in earlier
articles.

\textbf{Remark II}: The conditions 2. and 3. yield relationship between
(2.8) and (1.10). The inverse formulas are $p=\overset{\_}{p}/(1-\frac{%
\varepsilon }{2}\overset{\_}{p})$, \ \ $h=\overset{\_}{h}/(1-\frac{%
\varepsilon }{2}\overset{\_}{p})$, \ $a^{\alpha }=c^{\alpha }/(1-\frac{%
\varepsilon }{2}\overset{\_}{p})$.

\textbf{Proof}: An arbitrary conservation law for the evolutionary system
(1.7) can be presented in its divergent form $d\xi =hdx+fdt$. If we apply
the reciprocal transformation (3.4) for all ($N+2$) conservation laws
(1.7-9) and (1.11), then we at once obtain conditions of this theorem.

\section{Remarkable examples}

\begin{enumerate}
\item[1.] It is well-known fact (see for instance [7]) that \textbf{the
Calogero Korteweg-de Vries equation} (CKdV)
\end{enumerate}

\begin{equation}
u_{y}=\partial _{z}[u_{zz}+\frac{3}{2u}(1-u_{z}^{2})]  \tag{4.1}
\end{equation}
has the nonlocal Hamiltonian structure (1.14) (see (2.1))

\begin{equation}
u_{y}=[u^{2}\partial _{z}+uu_{z}-u_{z}\partial _{z}^{-1}u_{z}]\frac{\delta H%
}{\delta u}  \tag{4.2}
\end{equation}
where $H=-\frac{1}{2}\int \frac{1+u_{z}^{2}}{u^{2}}dz$. However, here $%
\overset{\_}{g}^{11}\equiv 0$. Thus, the CKdV equation has extraordinary
momentum $P=\int 1\cdot dz$ and two Casimirs $Q_{1}=\int udz$ and $%
Q_{2}=\int dz/u$. Here we introduce other particular reciprocal
transformation (see (3.2), (3.3) and (4.1))

\begin{equation}
dt=dy,\ \ \ dx=udz+[u_{zz}+\frac{3}{2u}(1-u_{z}^{2})]dy.  \tag{4.3}
\end{equation}
Then inverse reciprocal transformation is
\begin{equation}
dy=dt,\ \ \ dz=wdx+[\frac{w_{xx}}{w^{3}}-\frac{3w_{x}^{2}}{2w^{4}}-\frac{3}{2%
}w^{2}]dt.  \tag{4.4}
\end{equation}
where $uw=1$ and $\partial _{z}=u\partial _{x}$, and the CKdV equation
transforms into
\begin{equation}
w_{t}=\partial _{x}[\frac{w_{xx}}{w^{3}}-\frac{3w_{x}^{2}}{2w^{4}}-\frac{3}{2%
}w^{2}]  \tag{4.5}
\end{equation}
This equation (4.5) has local Hamiltonian structure
\begin{equation}
w_{t}=\partial _{x}\frac{\delta \overset{\_}{H}}{\delta w}  \tag{4.6}
\end{equation}
where ($\overset{\_}{H}=\int \overset{\_}{h}(w,w_{x})dx=\int \overset{\_}{h}%
(1/u,(1/u)_{z}/u)udz=\int h(u,u_{z})dz$) the Hamiltonian is $\overset{\_}{H}%
=-\frac{1}{2}\int [\frac{w_{x}^{2}}{w^{3}}+w^{3}]dx$, the Momentum is $%
\overset{\_}{P}=\frac{1}{2}Q_{2}=\frac{1}{2}\int dz/u=\frac{1}{2}\int
w^{2}dx $, the Casimir is $\overset{\_}{Q}=P=\int 1\cdot dz=\int wdx$ and
other Casimir for the nonlocal Hamiltonian structure (4.2) transforms into
trivial Casimir $Q_{1}=\int udz=\int 1\cdot dx$.

Here we introduce potential function $z$ (see (1.13) and (4.4)), then the
equation (4.5) has the Lagrangian representation
\begin{equation}
S=\frac{1}{2}\int [z_{x}z_{t}+\frac{z_{xx}^{2}}{z_{x}^{3}}+z_{x}^{3}]dxdt
\tag{4.7}
\end{equation}
where $w=z_{x}$. We can apply the reciprocal transformation (4.4) for the
2-form
\begin{equation*}
\Omega =[z_{x}z_{t}+\frac{z_{xx}^{2}}{z_{x}^{3}}+z_{x}^{3}]dx\wedge dt.
\end{equation*}
Then this 2-form
\begin{equation*}
\Omega =[we+\frac{w_{x}^{2}}{w^{3}}+w^{3}]dx\wedge dt,
\end{equation*}
where $e=z_{t}$ transforms into
\begin{equation*}
\Omega =[\frac{e}{u}+\frac{1+u_{z}^{2}}{u^{3}}]udz\wedge dy,
\end{equation*}
where $dx\wedge dt=udz\wedge dy$ and $\partial _{x}=w\partial _{z}$ (see
(4.3) and (4.4)). Thus, this 2-form
\begin{equation*}
\Omega =[-\frac{x_{y}}{x_{z}}+\frac{1+x_{zz}^{2}}{x_{z}^{2}}]dz\wedge dy,
\end{equation*}
where $u=x_{z}$ and $e=-x_{y}/x_{z}$ (see (4.3)) yields the Lagrangian
representation for the CKdV equation
\begin{equation}
S=\frac{1}{2}\int [-\frac{x_{y}}{x_{z}}+\frac{1+x_{zz}^{2}}{x_{z}^{2}}]dzdy,
\tag{4.8}
\end{equation}

Thus, we have described a relationship between a Lagrangian representation
for evolutionary equation with local Hamiltonian structure (4.6) and a
Lagrangian representation for evolutionary equation with nonlocal
Hamiltonian structure (4.2). This action (4.8) have been established in
article [8] for the Krichever-Novikov equation. However, here we will give a
generalization of this Lagrangian representation on $N$-component case for
evolutionary systems with nonlocal Hamiltonian structure (2.6).

\begin{enumerate}
\item[2.] \textbf{The Thrice-Modified Kaup-Boussinesq\ system} (see [9])
\end{enumerate}

\begin{equation}
c_{y}=\partial _{z}[-\frac{1}{2}b(1\text{+}b^{2})\text{+}\varepsilon
(cb_{z}-bc_{z})]\text{, \ }b_{y}=\partial _{z}[-\frac{(1\text{-}b^{2})^{2}}{%
2c}\text{+}\varepsilon (bb_{z}\text{+}\frac{1\text{-}b^{2}}{c}c_{z})]
\tag{4.9}
\end{equation}
($\varepsilon $ is arbitrary constant, not curvature here!) has nonlocal
\textit{Hamiltonian}\emph{\ }structure

\begin{eqnarray}
c_{y} &=&\frac{1}{4}\partial _{z}[bc\frac{\delta H}{\delta b}+c^{2}\frac{%
\delta H}{\delta c}-cF]  \TCItag{4.10} \\
b_{y} &=&\frac{1}{4}\partial _{z}[(b^{2}-1)\frac{\delta H}{\delta b}+bc\frac{%
\delta H}{\delta c}-bF],  \notag
\end{eqnarray}
where the Hamiltonian is $H=2\int [b(1-b^{2})-2\varepsilon bc_{z}]dz/c$ \
and $\partial _{z}F=\frac{\delta H}{\delta b}b_{z}+\frac{\delta H}{\delta c}%
c_{z}$. This bracket is determined by the differential-geometric \textit{%
Poisson}\emph{\ }bracket with metric of constant curvature (1.14)
\begin{eqnarray}
\{b(z),b(z^{\prime })\} &=&\frac{1}{4}[(b^{2}-1)\partial
_{z}+bb_{z}-b_{z}\partial _{z}^{-1}b_{z}]\,\delta (z-z^{\prime })
\TCItag{4.11} \\
\{b(z),c(z^{\prime })\} &=&\frac{1}{4}[cb\partial _{z}+cb_{z}-b_{z}\partial
_{z}^{-1}c_{z}]\,\delta (z-z^{\prime })  \notag \\
\{c(z),b(z^{\prime })\} &=&\frac{1}{4}[bc\partial _{z}+bc_{z}-c_{z}\partial
_{z}^{-1}b_{z}]\,\delta (z-z^{\prime })  \notag \\
\{c(z),c(z^{\prime })\} &=&\frac{1}{4}[c^{2}\partial
_{z}+cc_{z}-c_{z}\partial _{z}^{-1}c_{z}]\,\delta (z-z^{\prime }).  \notag
\end{eqnarray}
$\qquad \bigskip $\quad Since, just $\overset{\_}{g}^{11}=-1$ ($\overset{\_}{%
g}^{12}=\overset{\_}{g}^{21}=\overset{\_}{g}^{22}=0$, e.g. $\det \overset{\_}%
{g}^{\alpha \beta }=0$) the system (4.13) has extraordinary momentum $P=\int
1\cdot dz$ (compare with (2.4-2.6) and (2.8)), but three Casimirs

\begin{equation*}
Q_{1}=\int \frac{1-b^{2}}{c}dz,\qquad Q_{2}=\int bdz,\qquad Q_{3}=\int cdz.
\end{equation*}
These conservation law densities determine the constraint

\begin{equation}
q_{1}q_{3}+q_{2}^{2}=1  \tag{4.12}
\end{equation}
where $Q_{\alpha }=\int q_{\alpha }dz$, $\alpha =1,2,3$. The Poisson bracket
(4.11) can be reduced into canonical form (1.7) by multi-parameter
reciprocal transformation (3.2) (see (3.3)). Here we can for instance use
simplest particular reciprocal transformation$\qquad $

\begin{equation}
dt=dy,\quad dx=cdz+[-\frac{1}{2}b(1+b^{2})+\varepsilon (cb_{z}-bc_{z})]dy
\tag{4.13}
\end{equation}
(see the first equation in (4.9)).Then, the inverse reciprocal
transformation is

\begin{equation}
dy=dt,\quad dz=udx+\frac{1}{2}[w+\frac{w^{3}}{u^{2}}-2\varepsilon \frac{w_{x}%
}{u^{2}}]dt,  \tag{4.14}
\end{equation}
where $uc=1$, \ $b=wc$ and $\partial _{z}=c\partial _{x}$ or $\partial
_{x}=u\partial _{z}$. And we at once obtain other integrable system

\begin{equation}
u_{t}=\frac{1}{2}\partial _{x}[w+\frac{w^{3}}{u^{2}}-2\varepsilon \frac{w_{x}%
}{u^{2}}],\quad \quad w_{t}=\frac{1}{2}\partial _{x}[3\frac{w^{2}}{u}%
-u-2\varepsilon \frac{u_{x}}{u^{2}}],  \tag{4.15}
\end{equation}
which has local Hamiltonian structure\qquad

\begin{equation}
u_{t}=\frac{1}{4}\partial _{x}\frac{\delta \overset{\_}{H}}{\delta u},\qquad
w_{t}=-\frac{1}{4}\partial _{x}\frac{\delta \overset{\_}{H}}{\delta w}
\tag{4.16}
\end{equation}
with ($\overset{\_}{H}=\int \overset{\_}{h}(b,c,c_{z})dz=\int \overset{\_}{h}%
(w/u,1/u,(1/u)_{z}/u)udx=\int h(u,w,w_{x})dx$) the Hamiltonian $\overset{\_}{%
H}=2\int [uw-\frac{w^{3}-2\varepsilon w_{x}}{u}]dx$, also with Momentum $%
\overset{\_}{P}=2Q_{1}=2\int q_{1}dz=2\int (u^{2}-w^{2})dx$ (see (4.12)),
two Casimirs $\overset{\_}{Q}_{2}=Q_{2}=\int bdz=\int wdx$, \ $\overset{\_}{Q%
}_{3}=P=\int 1\cdot dz=\int udx$ and other Casimir for the nonlocal
Hamiltonians structure (4.10) transforms into trivial Casimir $Q_{3}=\int
cdz=\int 1\cdot dx$.

The evolutionary system (4.16) with the Poisson bracket

\begin{equation}
\{u\left( x\right) ,u\left( x^{\prime }\right) \}=-\{w\left( x\right)
,w\left( x^{\prime }\right) \}=\frac{1}{4}\delta ^{\prime }(x-x^{\prime })
\tag{4.17}
\end{equation}
has the Lagrangian representation (see (1.13) and (4.14))
\begin{equation}
S=\frac{1}{2}\int [z_{x}z_{t}-\varphi _{x}\varphi _{t}-\frac{2\varepsilon
\varphi _{xx}-\varphi _{x}^{3}}{z_{x}}-z_{x}\varphi _{x}]dxdt,  \tag{4.18}
\end{equation}
where $w=\varphi _{x}$. We can apply the reciprocal transformation (4.14)
for the 2-form
\begin{equation*}
\Omega =[z_{x}z_{t}-\varphi _{x}\varphi _{t}-\frac{2\varepsilon \varphi
_{xx}-\varphi _{x}^{3}}{z_{x}}-z_{x}\varphi _{x}]dx\wedge dt.
\end{equation*}
Then this 2-form
\begin{equation*}
\Omega =[ue-w\upsilon -\frac{2\varepsilon w_{x}-w^{3}}{u}-uw]dx\wedge dt,
\end{equation*}
where $e=z_{t}$ and $\upsilon =\varphi _{t}$, transforms itself into
\begin{equation*}
\Omega =[\frac{e}{c}-\frac{b\upsilon }{c}-2\varepsilon (\frac{b}{c})_{z}+%
\frac{b^{3}}{c^{2}}-\frac{b}{c^{2}}]cdz\wedge dy,
\end{equation*}
where $dx\wedge dt=cdz\wedge dy$ (see (4.13)). Since $d\varphi =wdx+\upsilon
dt=bdz+(\upsilon -be)dy$, then this 2-form
\begin{equation*}
\Omega =[-\frac{1-\varphi _{z}^{2}}{x_{z}}x_{y}-\varphi _{z}\varphi
_{y}-2\varepsilon x_{z}(\frac{\varphi _{z}}{x_{z}})_{z}+\frac{\varphi
_{z}^{3}}{x_{z}}-\frac{\varphi _{z}}{x_{z}}]dz\wedge dy,
\end{equation*}
where $c=x_{z}$, $e=-x_{y}/x_{z}$, $b=\varphi _{z}$ and $\upsilon =\varphi
_{y}-\varphi _{z}x_{y}/x_{z}$ (see (4.13)), yields the Lagrangian
representation for the Thrice-Modified Kaup-Boussinesq system
\begin{equation}
S=\frac{1}{2}\int [\frac{1-\varphi _{z}^{2}}{x_{z}}x_{y}+\varphi _{z}\varphi
_{y}-2\varepsilon \frac{x_{zz}}{x_{z}}\varphi _{z}+\frac{\varphi
_{z}(1-\varphi _{z}^{2})}{x_{z}}]dzdy.  \tag{4.19}
\end{equation}

\textbf{Remark}: The Kaup-Boussinesq system has nonlocal Hamiltonian
structure with degenerated constant $\overset{\_}{g}^{\alpha \beta }$matrix
(see (2.4)) in coordinates ($q_{2},q_{3}$) (see (4.12)). However, this
Poisson bracket (4.11) easily can be transformed into canonical form (2.6)
with canonical metrics (2.4), if we change variable $q_{2}\rightarrow 1-q$.
Then (4.12) at once yields a Momentum (see (2.8))
\begin{equation*}
q=1-\sqrt{1-q_{1}q_{3}}
\end{equation*}
expressed in its canonical variables Casimirs $q_{1}$ and $q_{3}$ (it is
easy to see, that the Momentum $P=\int qdz$ (where $q=1-q_{2}$) is a linear
combination of two Casimirs $Q_{3}$ and $Q_{2}$.). Immediately we obtain
(see (2.8)) all non-zero components of constant matrix $\overset{\_}{g}%
_{\alpha \beta }$: $\overset{\_}{g}_{12}=\overset{\_}{g}_{21}=1/2$ and
curvature $\varepsilon =1$. Thus, $\overset{\_}{g}^{12}=\overset{\_}{g}%
^{21}=2$ and the Thrice-Modified Kaup-Boussinesq system (4.9) can be
re-written in variables ($r,c$) (see (2.4) and (2.6)), where $rc+b^{2}=1$
(see (4.12))
\begin{equation*}
r_{y}=\frac{1}{2}\partial _{z}[\frac{r^{2}}{\sqrt{1\text{-}rc}}\text{+}%
\varepsilon \frac{r^{2}c_{z}\text{+}(2\text{-}rc)r_{z}}{\sqrt{1-rc}}]\text{,
\ }c_{y}=-\frac{1}{2}\partial _{z}[(2\text{-}rc)\sqrt{1\text{-}rc}%
+\varepsilon \frac{(2\text{-}rc)c_{z}\text{+}c^{2}r_{z}}{\sqrt{1-rc}}].
\end{equation*}
This system has nonlocal Hamiltonian structure (2.6)
\begin{equation*}
r_{y}=\frac{1}{4}\partial _{z}[-r^{2}\frac{\delta H}{\delta r}+(2-rc)\frac{%
\delta H}{\delta c}+rF]\text{, \ }c_{y}=\frac{1}{4}\partial _{z}[(2-rc)\frac{%
\delta H}{\delta r}-c^{2}\frac{\delta H}{\delta c}+cF],
\end{equation*}
where $H=2\int \sqrt{1-rc}(r-2\varepsilon c_{z}/c)dz$ and $\partial _{z}F=%
\frac{\delta H}{\delta r}r_{z}+\frac{\delta H}{\delta c}c_{z}$ (see (4.10)).
Thus, the Poisson bracket (4.11) determined by Lagrangian representation
(4.19) for the evolutionary system (4.9) yields the canonical Poisson
bracket (2.4)

\begin{eqnarray*}
\{r(z),r(z^{\prime })\} &=&\frac{1}{4}[-r^{2}\partial
_{z}-rr_{z}+r_{z}\partial _{z}^{-1}r_{z}]\,\delta (z-z^{\prime }), \\
\{r(z),c(z^{\prime })\} &=&\frac{1}{4}[(2-rc)\partial
_{z}-cr_{z}+r_{z}\partial _{z}^{-1}c_{z}]\,\delta (z-z^{\prime }), \\
\{c(z),r(z^{\prime })\} &=&\frac{1}{4}[(2-rc)\partial
_{z}-rc_{z}+c_{z}\partial _{z}^{-1}r_{z}]\,\delta (z-z^{\prime }), \\
\{c(z),c(z^{\prime })\} &=&\frac{1}{4}[-c^{2}\partial
_{z}-cc_{z}+c_{z}\partial _{z}^{-1}c_{z}]\,\delta (z-z^{\prime }).
\end{eqnarray*}

\section{Lagrangian Representation}

\textbf{Major result}:

\begin{theorem}
The evolutionary system (1.1) with nonlocal Hamiltonian structure (see
(2.1)) determined by differential-geometric Poisson bracket of the first
order associated with metrics of constant curvature (1.14) has the
Lagrangian representation
\begin{equation}
S=\int [\frac{1-\overset{\_}{g}_{\alpha \nu }\varphi _{z}^{\alpha }\varphi
_{z}^{\nu }}{2x_{z}}x_{y}+\frac{1}{2}\overset{\_}{g}_{\alpha \nu }\varphi
_{z}^{\alpha }\varphi _{y}^{\nu }-h(x_{z},\mathbf{\varphi }_{z},x_{zz},%
\mathbf{\varphi }_{zz},...)]dzdy  \tag{5.1}
\end{equation}
\end{theorem}

\textbf{Proof}: From variational derivatives $\ \delta S/\delta \varphi
^{\alpha }=0$, $\ \delta S/\delta x=0$ \ and compatibility condition $%
(x_{z})_{y}=(x_{y})_{z}$ , respectively, we obtain evolutionary system on ($%
N+2$) equations
\begin{equation}
\upsilon _{y}^{\alpha }\text{+}\partial _{z}(\rho \upsilon ^{\alpha }-%
\overset{\_}{g}^{\alpha \nu }\delta H/\delta \upsilon ^{\nu })=0,\ w_{y}%
\text{+}\partial _{z}(\rho w-\delta H/\delta u)=0,\ u_{y}\text{+}\partial
_{z}(\rho u)=0,  \tag{5.2}
\end{equation}
where $u=x_{z}$, \ $\rho =-x_{y}/x_{z}$, \ $\upsilon ^{\alpha }=\varphi
_{z}^{\alpha }$, \ $H=\int h(u,\mathbf{\upsilon },u_{z},\mathbf{\upsilon }%
_{z},...)dz$ and constraint
\begin{equation}
1=2uw+\overset{\_}{g}_{\alpha \nu }\upsilon ^{\alpha }\upsilon ^{\nu }.
\tag{5.3}
\end{equation}
The system (5.2) is over-determined system. Thus, from the obvious condition
$\partial _{y}(2uw+\overset{\_}{g}_{\alpha \nu }\upsilon ^{\alpha }\upsilon
^{\nu })=0$ (see (5.3)) we obtain an explicit expression of the function $%
\rho $

\begin{equation}
\rho =u\frac{\delta H}{\delta u}+\upsilon ^{\alpha }\frac{\delta H}{\delta
\upsilon ^{\alpha }}-F,  \tag{5.4}
\end{equation}
where $\partial _{z}F=\frac{\delta H}{\delta \upsilon ^{\alpha }}\upsilon
_{z}^{\alpha }+\frac{\delta H}{\delta u}u_{z}$ (see (1.12)).

At first we introduce new variables ($p,q,\mathbf{\upsilon }$) by

\begin{equation}
u=1-p-q,\ \ 2w=1-p+q,\ \ \ q^{\alpha }=\upsilon ^{\alpha },  \tag{5.5}
\end{equation}
then all partial derivatives are
\begin{equation}
\frac{\partial h}{\partial u}=-\frac{1-p}{1-p-q}\frac{\partial h}{\partial q}%
\text{, \ \ \ }\frac{\partial h}{\partial \upsilon ^{\alpha }}=\frac{%
\partial h}{\partial q^{\alpha }}-\frac{\overset{\_}{g}_{\alpha \nu }q^{\nu }%
}{1-p-q}\frac{\partial h}{\partial q}.  \tag{5.6}
\end{equation}
For simplicity and without loss of generality it is sufficient, if we will
study just the hydrodynamic type case, where $H=\int h(u,\upsilon )dz$. Then
(see (5.4))
\begin{equation}
\rho =(q\frac{\partial h}{\partial q}+q^{\alpha }\frac{\partial h}{\partial
q^{\alpha }}-h)-\frac{1}{1-p-q}\frac{\partial h}{\partial q}  \tag{5.7}
\end{equation}
and system (5.2) transforms itself into
\begin{equation}
q_{y}\text{=}\partial _{z}[-\frac{\partial h}{\partial q}-q(q\frac{\partial h%
}{\partial q}\text{+}q^{\alpha }\frac{\partial h}{\partial q^{\alpha }}\text{%
-}h)]\text{, \ }q_{y}^{\alpha }\text{=}\partial _{z}[\overset{\_}{g}^{\alpha
\nu }\frac{\partial h}{\partial q^{\nu }}-q^{\alpha }(q\frac{\partial h}{%
\partial q}\text{+}q^{\nu }\frac{\partial h}{\partial q^{\nu }}\text{-}h)].
\tag{5.8}
\end{equation}
This system has the momentum (see (5.3) and (5.5))
\begin{equation}
p=1-\sqrt{1+q^{2}-\overset{\_}{g}_{\alpha \nu }q^{\alpha }q^{\nu }}
\tag{5.9}
\end{equation}
(thus, curvature $\varepsilon =1$, see (2.8)) and conservation law of
momentum is
\begin{equation}
p_{y}=\partial _{z}[(1-p)(q\frac{\partial h}{\partial q}+q^{\alpha }\frac{%
\partial h}{\partial q^{\alpha }}-h)].  \tag{5.10}
\end{equation}
Here we introduce new variables $\ c^{0}=q$, $\ c^{\alpha }=q^{\alpha }$, $\
\ \alpha =1,2...N$. Then the system (5.8) is exactly the system (2.6) with
constant matrices

\begin{equation}
\overset{\symbol{126}}{g}^{\alpha \nu }=\left(
\begin{array}{cc}
-1 & 0 \\
0 & \overset{\_}{g}^{\alpha \nu }%
\end{array}
\right) \text{, \ \ \ }\overset{\symbol{126}}{g}_{\alpha \nu }=\left(
\begin{array}{cc}
-1 & 0 \\
0 & \overset{\_}{g}_{\alpha \nu }%
\end{array}
\right) .  \tag{5.11}
\end{equation}

\textbf{Remark}: If we introduce the reciprocal transformation (see (5.2))
\begin{equation}
dt=dy\text{, \ \ \ }dx=udz-\rho udy  \tag{5.12}
\end{equation}
we can apply (5.12) for 2-form (see (5.1) and Section IV)
\begin{equation*}
\Omega =[-\rho \frac{1-\overset{\_}{g}_{\alpha \nu }\upsilon ^{\alpha
}\upsilon ^{\nu }}{2}+\frac{1}{2}\overset{\_}{g}_{\alpha \nu }\upsilon
^{\alpha }e^{\nu }-h(u,\mathbf{\upsilon },u_{z},\mathbf{\upsilon }%
_{z},...)]dz\wedge dy,
\end{equation*}
where $e^{\alpha }=\varphi _{y}^{\alpha }$. Then this 2-form (where $%
dx\wedge dt=udz\wedge dy$, see (5.12))
\begin{equation*}
\Omega =[-z_{t}\frac{z_{x}^{2}-\overset{\_}{g}_{\alpha \nu }\varphi
_{x}^{\alpha }\varphi _{x}^{\nu }}{2z_{x}}+\frac{1}{2z_{x}}\overset{\_}{g}%
_{\alpha \nu }\varphi _{x}^{\alpha }(\varphi _{t}^{\nu }z_{x}-\varphi
_{x}^{\nu }z_{t})-z_{x}h]dx\wedge dt,
\end{equation*}
where $z_{x}=1/u$, \ $z_{t}=\rho $, \ \ $\upsilon ^{\alpha }=\varphi
_{x}^{\alpha }/z_{x}$, \ \ $e^{\alpha }=\varphi _{t}^{\alpha }-z_{t}\varphi
_{x}^{\alpha }/z_{x}$, determines the action (see (1.13))
\begin{equation}
S=\int [-\frac{1}{2}z_{x}z_{t}+\frac{1}{2}\overset{\_}{g}_{\alpha \nu
}\varphi _{x}^{\alpha }\varphi _{t}^{\nu }-\overset{\_}{h}(z_{x},\mathbf{%
\varphi }_{x},z_{xx},\mathbf{\varphi }_{xx},...)]dxdt  \tag{5.12}
\end{equation}
for the evolutionary system (1.1) with local Hamiltonian structure (1.7) and
constant metrics (5.11), where (see (1.7)) $a^{0}=z_{x}$, $a^{\alpha
}=\varphi _{x}^{\alpha }$ ($\alpha =1,2...N$) and $\overset{\_}{h}=a^{0}h$ ($%
H=\int hdz=\int \overset{\_}{h}dx=\int a^{0}hdx$).

\section{Kaup-Boussinesq system and its nonlocal Hamiltonian structure}

Many different integrable systems have different local and nonlocal
Hamiltonian structures. For example the Korteweg-de Vries equation has two
local Hamiltonian structures and all others are nonlocal. The
Kaup-Boussinesq system (see for instance [9])

\begin{equation}
\upsilon _{y}=\partial _{z}[\frac{1}{2}\upsilon ^{2}+\eta ],\qquad \eta
_{y}=\partial _{z}[\upsilon \eta +\varepsilon ^{2}\upsilon _{zz}],  \tag{6.1}
\end{equation}
has the\ three local Hamiltonian structures determined by following Poisson
brackets
\begin{eqnarray}
\{\upsilon ,\eta \}_{1} &=&\{\eta ,\upsilon \}_{1}=\delta ^{\prime
}(z-z^{\prime }),  \TCItag{6.2} \\
&&  \notag \\
\{\upsilon ,\upsilon \}_{2} &=&\delta ^{\prime }(z-z^{\prime })\text{, \ \ }%
\{\upsilon ,\eta \}_{2}=\frac{1}{2}(\upsilon \partial _{z}+\upsilon
_{z})\delta (z-z^{\prime }),  \TCItag{6.3} \\
\{\eta ,\upsilon \}_{2} &=&\frac{1}{2}\upsilon \delta ^{\prime }(z-z^{\prime
})\text{, \ \ }\{\eta ,\eta \}_{2}=\varepsilon ^{2}\delta ^{\prime \prime
\prime }(z-z^{\prime })+(\eta \partial _{z}+\frac{1}{2}\eta _{z})\delta
(z-z^{\prime }),  \notag \\
&&  \notag \\
\{\upsilon ,\upsilon \}_{3} &=&(\upsilon \partial _{z}+\frac{1}{2}\upsilon
_{z})\delta (z-z^{\prime }),  \TCItag{6.4} \\
\{\upsilon ,\eta \}_{3} &=&\varepsilon ^{2}\delta ^{\prime \prime \prime
}(z-z^{\prime })+\frac{1}{4}[(\upsilon ^{2}+4\eta )\partial _{z}+(\upsilon
^{2}+2\eta )_{z}]\delta (z-z^{\prime }),  \notag \\
\{\eta ,\upsilon \}_{3} &=&\varepsilon ^{2}\delta ^{\prime \prime \prime
}(z-z^{\prime })+\frac{1}{4}[(\upsilon ^{2}+4\eta )\partial _{z}+2\eta
_{z}]\delta (z-z^{\prime }),  \notag \\
\{\eta ,\eta \}_{3} &=&\frac{\varepsilon ^{2}}{2}[2\upsilon \partial _{z}^{3}%
\text{+}3\upsilon _{z}\partial _{z}^{2}\text{+}3\upsilon _{zz}\partial _{z}%
\text{+}\upsilon _{zzz}]\delta (z-z^{\prime })\text{+}[\upsilon \eta
\partial _{z}\text{+}\frac{1}{2}(\upsilon \eta )_{z}]\delta (z-z^{\prime })
\notag
\end{eqnarray}
and nonlocal Hamiltonian structures, where the first of them is
\begin{eqnarray}
\{\upsilon ,\upsilon \}_{4} &=&\varepsilon ^{2}\delta ^{\prime \prime \prime
}(z-z^{\prime })\text{+}\frac{1}{4}[(3\upsilon ^{2}+4\eta )\partial _{z}%
\text{+}\frac{1}{2}(3\upsilon ^{2}+4\eta )_{z}-\upsilon _{z}\partial
_{z}^{-1}\upsilon _{z}]\delta (z-z^{\prime }),  \notag \\
\{\upsilon ,\eta \}_{4} &=&\frac{\varepsilon ^{2}}{2}[3\upsilon \partial
_{z}^{3}+4\upsilon _{z}\partial _{z}^{2}+3\upsilon _{zz}\partial
_{z}+\upsilon _{zzz}]\delta (z-z^{\prime })+\frac{1}{4}[(6\upsilon \eta +%
\frac{1}{2}\upsilon ^{3})\partial _{z}+  \notag \\
&&+(\frac{3}{2}\upsilon ^{2}\upsilon _{z}+4\eta \upsilon _{z}+3\upsilon \eta
_{z})-\upsilon _{z}\partial _{z}^{-1}\eta _{z}]\delta (z-z^{\prime }),
\TCItag{6.5} \\
\{\eta ,\upsilon \}_{4} &=&\frac{\varepsilon ^{2}}{2}[3\upsilon \partial
_{z}^{3}+5\upsilon _{z}\partial _{z}^{2}+4\upsilon _{zz}\partial
_{z}+\upsilon _{zzz}]\delta (z-z^{\prime })+\frac{1}{4}[(6\upsilon \eta +%
\frac{1}{2}\upsilon ^{3})\partial _{z}+  \notag \\
&&+(2\eta \upsilon _{z}+3\upsilon \eta _{z})-\eta _{z}\partial
_{z}^{-1}\upsilon _{z}]\delta (z-z^{\prime }),  \notag \\
\{\eta ,\eta \}_{4} &=&\varepsilon ^{4}\delta ^{V}(z-z^{\prime })+\frac{%
\varepsilon ^{2}}{4}[(8\eta +3\upsilon ^{2})\partial _{z}^{3}+\frac{3}{2}%
(8\eta +3\upsilon ^{2})_{z}\partial _{z}^{2}+  \notag \\
&&[(8\eta +3\upsilon ^{2})_{zz}+3\upsilon \upsilon _{zz}]\partial
_{z}+[(2\eta +\upsilon ^{2})_{zz}+\upsilon \upsilon _{zz}-\frac{1}{2}%
\upsilon _{z}^{2}]_{z}]\delta (z-z^{\prime })+  \notag \\
&&\frac{1}{4}[(4\eta ^{2}+3\upsilon ^{2}\eta )\partial _{z}+\frac{1}{2}%
(4\eta ^{2}+3\upsilon ^{2}\eta )_{z}-\eta _{z}\partial _{z}^{-1}\eta
_{z}]\delta (z-z^{\prime }).  \notag
\end{eqnarray}
The first Miura transformation
\begin{equation}
\eta =(\upsilon ^{2}-a^{2})/4-\varepsilon a_{z}  \tag{6.6}
\end{equation}
connects the Kaup-Boussinesq system (6.1) and the Modified Kaup-Boussinesq
system
\begin{equation}
a_{y}=\partial _{z}[\frac{1}{2}\upsilon a-\varepsilon \upsilon _{z}],\quad
\upsilon _{y}=\partial _{z}[\frac{1}{4}(3\upsilon ^{2}-a^{2})-\varepsilon
a_{z}],  \tag{6.7}
\end{equation}
which has two local Hamiltonian structures determined by the Poisson brackets

\begin{eqnarray}
\{a,a\}_{1} &=&-\delta ^{\prime }(z-z^{\prime }),\quad \{\upsilon ,\upsilon
\}_{1}=\delta ^{\prime }(z-z^{\prime }),  \TCItag{6.8} \\
&&  \notag \\
\{a,a\}_{2} &=&0,\quad \{a,\upsilon \}_{2}=-\varepsilon \delta ^{\prime
\prime }(z-z^{\prime })+\frac{1}{2}(a\partial _{z}+a_{z})\delta (z-z^{\prime
}),  \TCItag{6.9} \\
\{\upsilon ,a\}_{2} &=&\varepsilon \delta ^{\prime \prime }(z-z^{\prime })+%
\frac{1}{2}a\delta ^{\prime }(z-z^{\prime })\text{, \ \ }\{\upsilon
,\upsilon \}_{2}=(\upsilon \partial _{z}+\frac{1}{2}\upsilon _{z})\delta
(z-z^{\prime })  \notag
\end{eqnarray}
and nonlocal Hamiltonian structures, where the first of them is
\begin{eqnarray}
\{a,a\}_{3} &=&-\varepsilon ^{2}\delta ^{\prime \prime \prime }(z-z^{\prime
})+\frac{1}{4}[a^{2}\partial _{z}+aa_{z}-a_{z}\partial _{z}^{-1}a_{z}]\delta
(z-z^{\prime }),  \TCItag{6.10} \\
\{a,\upsilon \}_{3}\text{{}} &=&\text{{}}[-\frac{\varepsilon }{2}(2\upsilon
\partial _{z}^{2}\text{+}3\upsilon _{z}\partial _{z}\text{+}\upsilon _{zz})%
\text{+}\frac{1}{4}(2a\upsilon \partial _{z}\text{+}a\upsilon _{z}\text{+}%
2\upsilon a_{z}-a_{z}\partial _{z}^{-1}\upsilon _{z})]\delta (z\text{-}%
z^{\prime }),  \notag \\
\{\upsilon ,a\}_{3} &=&\varepsilon (\upsilon \partial _{z}+\frac{1}{2}%
\upsilon _{z})\delta ^{\prime }(z-z^{\prime })+\frac{1}{4}[2a\upsilon
\partial _{z}+a\upsilon _{z}-\upsilon _{z}\partial _{z}^{-1}a_{z}]\delta
(z-z^{\prime }),  \notag \\
\{\upsilon ,\upsilon \}_{3} &=&\varepsilon ^{2}\delta ^{\prime \prime \prime
}(z-z^{\prime })\text{+}\frac{1}{4}[(3\upsilon ^{2}\text{+}4\eta )\partial
_{z}\text{+}\frac{1}{2}(3\upsilon ^{2}\text{+}4\eta )_{z}-\upsilon
_{z}\partial _{z}^{-1}\upsilon _{z}]\delta (z-z^{\prime }).  \notag
\end{eqnarray}
The second Miura transformation
\begin{equation}
\upsilon =a\,b+2\varepsilon b_{z}  \tag{6.11}
\end{equation}
connects the Modified Kaup-Boussinesq system (6.7) and the Twice-Modified
Kaup-Boussinesq system
\begin{equation}
b_{y}=\frac{1}{2}\partial _{z}[a(b^{2}-1)+2\varepsilon bb_{z}]\text{, \quad }%
a_{y}=\frac{1}{2}\partial _{z}[a^{2}b-2\varepsilon ba_{z}-4\varepsilon
^{2}b_{zz}],  \tag{6.12}
\end{equation}
which has one local Hamiltonian structure determined by the Poisson bracket
\begin{equation}
\{b,a\}_{1}=\frac{1}{2}\,\delta ^{\prime }(z-z^{\prime }),\quad \{a,b\}_{1}=%
\frac{1}{2}\delta ^{\prime }(z-z^{\prime })  \tag{6.13}
\end{equation}
and nonlocal Hamiltonian structures, where the first of them is
\begin{eqnarray}
\{b,b\}_{2} &=&\frac{1}{4}[(b^{2}-1)\partial _{z}+bb_{z}-b_{z}\partial
_{z}^{-1}b_{z}]\delta (z-z^{\prime }),  \TCItag{6.14} \\
\{b,a\}_{2} &=&\frac{\varepsilon }{2}(b\partial _{z}+b_{z})\delta ^{\prime
}(z-z^{\prime })+\frac{1}{4}[ab\partial _{z}+ab_{z}-b_{z}\partial
_{z}^{-1}a_{z}]\delta (z-z^{\prime }),  \notag \\
\{a,b\}_{2} &=&-\frac{\varepsilon }{2}(b\partial _{z}+b_{z})\delta ^{\prime
}(z-z^{\prime })+\frac{1}{4}[ab\partial _{z}+ba_{z}-a_{z}\partial
_{z}^{-1}b_{z}]\delta (z-z^{\prime }),  \notag \\
\{a,a\}_{2} &=&-\varepsilon ^{2}\delta ^{\prime \prime \prime }(z-z^{\prime
})+\frac{1}{4}[a^{2}\partial _{z}+aa_{z}-a_{z}\partial _{z}^{-1}a_{z}]\delta
(z-z^{\prime }),  \notag
\end{eqnarray}
The third Miura transformation (see for comparison (4.12))
\begin{equation}
ac+b^{2}+2\varepsilon c_{z}=1  \tag{6.15}
\end{equation}
connects the Twice-Modified Kaup-Boussinesq system (6.12) and the
Thrice-Modified Kaup-Boussinesq system (4.9) which has just nonlocal
Hamiltonian structures, where the first of them is determined by the Poisson
bracket (4.11).

Thus, the second local Hamiltonian structure (see (6.3)) of the
Kaup-Boussinesq system (6.1) is the first local Hamiltonian structure (see
(6.8)) of the Modified Kaup-Boussinesq system (6.7). The third local
Hamiltonian structure (see (6.4)) of Kaup-Boussinesq system (6.1) is the
second local Hamiltonian structure (see (6.9)) of the Modified
Kaup-Boussinesq system (6.7), which is the first local Hamiltonian structure
(see (6.13)) of the Twice-Modified Kaup-Boussinesq system (6.12). Moreover,
the Kaup-Boussinesq system (6.1) has fourth nonlocal Hamiltonian structure
(see (6.5)), which is the third nonlocal Hamiltonian structure (see (6.10))
of the Modified Kaup-Boussinesq system (6.7), also which is the second
nonlocal Hamiltonian structure (see (6.14)) of the Twice-Modified
Kaup-Boussinesq system (6.12), as well which is the first nonlocal
Hamiltonian structure (4.10) (see (4.11)) of the Thrice-Modified
Kaup-Boussinesq system (4.9) (see [9]). Thus, the Kaup-Boussinesq system
(6.1) has four different Lagrangian representations
\begin{eqnarray}
S_{1} &=&\int [\frac{1}{2}(\psi _{z}^{(1)}\psi _{y}^{(2)}+\psi
_{z}^{(2)}\psi _{y}^{(1)})-h_{4}(\psi _{z}^{(1)},\psi _{z}^{(2)},\psi
_{zz}^{(2)})]dzdy,  \TCItag{6.15} \\
S_{2} &=&\int [-\frac{1}{2}\psi _{z}^{(2)}\psi _{y}^{(2)}+\frac{1}{8}\psi
_{z}^{(3)}\psi _{y}^{(3)}-h_{3}(\psi _{z}^{(2)},\psi _{z}^{(3)},\psi
_{zz}^{(3)})]dzdy,  \TCItag{6.16} \\
S_{3} &=&\int [\frac{1}{2}(\psi _{z}^{(3)}\psi _{y}^{(4)}+\psi
_{z}^{(4)}\psi _{y}^{(3)})-h_{2}(\psi _{z}^{(3)},\psi _{z}^{(4)},\psi
_{zz}^{(4)})]dzdy,  \TCItag{6.17} \\
S_{4} &=&\int [\frac{1-\psi _{z}^{(4)}{}^{2}}{2\psi _{z}^{(5)}}\psi
_{y}^{(5)}+\frac{1}{2}\psi _{z}^{(4)}\psi _{y}^{(4)}-h_{1}(\psi
_{z}^{(4)},\psi _{z}^{(5)},\psi _{zz}^{(5)})]dzdy,  \TCItag{6.18}
\end{eqnarray}
where $\eta =\psi _{z}^{(1)}$, $\upsilon =\psi _{z}^{(2)}$, $a=\psi
_{z}^{(3)}$, $b=\psi _{z}^{(4)}$, $c=\psi _{z}^{(5)}$ and $H_{k}=\int
h_{k}\,dz$. Moreover, we have very interesting hierarchy:

\begin{enumerate}
\item The first Hamiltonian structure (see (6.2)) of the Kaup-Boussinesq
system (6.1) has the Hamiltonian $H_{4}=\frac{1}{2}\int [-\varepsilon
^{2}\upsilon _{z}^{2}+\upsilon ^{2}\eta +\eta ^{2}]\,dz$, the momentum $%
H_{3}=\int \upsilon \eta dz$ and two flat coordinates (Casimirs) $%
H_{2}=2\int \eta dz$, $H_{1}=2\int \upsilon dz$.

\item The first Hamiltonian structure (see (6.8)) of the Modified
Kaup-Boussinesq system (6.7) has the Hamiltonian $H_{3}=\int \upsilon \eta
dz=\frac{1}{4}\int [\upsilon (\upsilon ^{2}-a^{2})-4\varepsilon \upsilon
a_{z}]dz$, the momentum $H_{2}=2\int \eta dz=\frac{1}{2}\int [\upsilon
^{2}-a^{2}]dz$ and two flat coordinates $H_{1}=2\int \upsilon dz$, $%
H_{-1}=\int adz$.

\item The first Hamiltonian structure (see (6.13)) of the Twice-Modified
Kaup-Boussinesq system (6.12) has the Hamiltonian $H_{2}=\frac{1}{2}\int
[\upsilon ^{2}-a^{2}]dz=\int [-\frac{1}{2}a^{2}(1-b^{2})+2\varepsilon
abb_{z}+2\varepsilon ^{2}b_{z}^{2}]dz$, the momentum

$H_{1}=2\int \upsilon dz=2\int abdz$ and two flat coordinates $H_{-1}=\int
adz$, $H_{-2}=\int bdz$.

\item The first Hamiltonian structure (4.10) of the Thrice-Modified
Kaup-Boussinesq system (4.9) has the Hamiltonian $H_{1}=2\int abdz=2\int
[b(1-b^{2})-2\varepsilon bc_{z}]dz/c$, the momentum $H_{0}=\int 1\cdot dz$
and two ''geodesic'' coordinates $H_{-2}=\int bdz$, $H_{-3}=\int cdz$.

The generalization of the Darboux theorem on infinite-dimensional case
signifies that every (local or nonlocal) Hamiltonian structure of integrable
system can be reduced into canonical form ''d/dx''. For instance, it means
that every Hamiltonian structure of the Kaup-Boussinesq system possesses a
Lagrangian representation (see (6.15-18)).
\end{enumerate}

Thus, here we present canonical representation for the first four
Hamiltonian structures of the Kaup-Boussinesq system (6.1) (see above)
\begin{eqnarray*}
\upsilon _{y} &=&\partial _{z}\frac{\delta H_{4}}{\delta \eta }\text{, \ \ }%
\ \ \eta _{y}=\partial _{z}\frac{\delta H_{4}}{\delta \upsilon }, \\
a_{y} &=&-\partial _{z}\frac{\delta H_{3}}{\delta a}\text{, }\ \ \upsilon
_{y}=\partial _{z}\frac{\delta H_{3}}{\delta \upsilon }, \\
b_{y} &=&\frac{1}{2}\partial _{z}\frac{\delta H_{2}}{\delta a},\ \ a_{y}=%
\frac{1}{2}\partial _{z}\frac{\delta H_{2}}{\delta b}, \\
u_{t} &=&\frac{1}{4}\partial _{x}\frac{\delta H_{1}}{\delta u},\ \ w_{t}=-%
\frac{1}{4}\partial _{x}\frac{\delta H_{1}}{\delta w},
\end{eqnarray*}
which are determined by the Poisson brackets (6.2), (6.8), (6.13) and
(4.16), respectively.\bigskip

\textbf{Conclusion.\medskip }

In this article we established the Lagrangian representation for an
evolutionary system, where a nonlocal Hamiltonian structure is determined by
the differential-geometric Poisson bracket of the first order with metric of
constant curvature. Also, we presented canonical coordinates for the first
four Hamiltonian structures of the Kaup-Boussinesq system, where every of
them is in canonical form ''d/dx'' with a Lagrangian representation. In
theory of Hamiltonian structures for dispersive systems just two
differential-geometric Poisson brackets of first order allow special
coordinates (annihilators), where they are the exactly the same as for
hydrodynamic type systems. It means that: if anyone start from Poisson
bracket
\begin{equation}
\{u^{i}(x),u^{k}(x^{\prime })\}=[a_{0}^{ik}\partial
_{x}^{N}+a_{1}^{ik}\partial _{x}^{N-1}+...+a_{N}^{ik}]\delta (x-x^{\prime }),
\tag{c.1}
\end{equation}
where all functions ($a_{j}^{ik}$) are functions with respect to field
variables $u^{i}$ and their derivatives, then in some cases by special
differential substitutions this expression may be transform into canonical
(see (1.7) and above first three local Hamiltonian structures for
Kaup-Boussinesq system, first two local Hamiltonian structures for Modified
Kaup-Boussinesq system, first local Hamiltonian structure for Twice-Modified
Kaup-Boussinesq system)
\begin{equation}
\{a^{\alpha }(x),a^{\beta }(x^{\prime })\}=\overset{\_}{g}^{\alpha \beta
}\delta ^{\prime }(x-x^{\prime }).  \tag{c.1a}
\end{equation}
It means that Poisson bracket determine Hamiltonian structure for dispersive
system reducible to canonical form ''d/dx''. If anyone start from Poisson
bracket
\begin{equation}
\{u^{i}(x),u^{k}(x^{\prime })\}=[a_{0}^{ik}\partial
_{x}^{N}+a_{1}^{ik}\partial _{x}^{N-1}+...+a_{N}^{ik}+\varepsilon
u_{x}^{i}\partial _{x}^{-1}u_{x}^{k},  \tag{c.2}
\end{equation}
where all functions ($a_{j}^{ik}$) are functions with respect to field
variables $u^{i}$ and their derivatives, then in some cases by special
differential substitutions this expression may be transform into canonical
Poisson bracket associated with metric of constant curvature (see (1.14),
fourth Hamiltonian structure for Kaup-Boussinesq system, third Hamiltonian
structure for Modified Kaup-Boussinesq system, second Hamiltonian structure
for the Twice-Modified Kaup-Boussinesq system and first Hamiltonian
structure for the Thrice-Modified Kaup-Boussinesq system)
\begin{equation}
\{c^{\alpha }(x),c^{\beta }(x^{\prime })\}=[(\overset{\_}{g}^{\alpha \beta
}-\varepsilon c^{\alpha }c^{\beta })\partial _{x}-\varepsilon c^{\beta
}c_{x}^{\alpha }+\varepsilon c_{x}^{\alpha }\partial _{x}^{-1}c_{x}^{\beta
}]\delta (x-x^{\prime }).  \tag{c.2a}
\end{equation}
If anyone start from Poisson bracket
\begin{equation*}
\{u^{i}(x),u^{k}(x^{\prime })\}=[a_{0}^{ik}\partial
_{x}^{N}+a_{1}^{ik}\partial _{x}^{N-1}+...+a_{N}^{ik}+\varepsilon _{\alpha
\beta }w^{\alpha ,i}\partial _{x}^{-1}w^{\beta ,k},
\end{equation*}
where all functions ($a_{j}^{ik}$, $w^{\alpha ,i}$) are functions with
respect to field variables $u^{i}$ and their derivatives, then \textbf{not
exist any differential substitutions possessing reduction to
differential-geometric nonlocal Poisson bracket of the first order} (see
[10]). Thus, two Poisson brackets of arbitrary order (see (c.1 and c.2) may
be reduced into differential geometric Poisson brackets of the first order
arising in theory of hydrodynamic type systems (see [2], [4] and [10]). In
next article, we will describe the infinite-dimensional analogue of Darboux
theorem for all other nonlocal Hamiltonian structures for the
Kaup-Boussinesq system and we will construct their Lagrangian
representations. In this case we shall describe all types of nonlocal
Hamiltonian structures where corresponding symplectic structures are local.
Theory of more complicated nonlocal Hamiltonian structures were established
in [10] and [11]. Our statement is that every nonlocal Hamiltonian structure
determined by the differential-geometric Poisson bracket of the first order
(see [10]) has a Lagrangian representation. It means that every integrable
system like Kaup-Boussinesq system has an infinite set of Lagrangian
representations (see for instance [12]).\bigskip

\textbf{Acknowledgment. \medskip }

I would like to thank Dr. Eugenie Ferapontov and Prof. Serguey Tsarev for
their useful comments. This work is partially supported by a Russian Fund of
Scientific Research (00-01-00210 \U{438} 00-01-00366).\bigskip

\textbf{References:\medskip }

[\textbf{1]}. L.D.Faddeev, V.E.Zakharov, The Korteweg-de Vries equation is a

\ \ \ \ \ \ fully\ integrable Hamiltonian system, (Russian), Functional
Analysis

\ \ \ \ \ \ and its Appl. 5 (1971), No.4, 18-27.

[\textbf{2]}. B.A.Dubrovin, S.P.Novikov, Hamiltonian formalism for
one-dimensional

\ \ \ \ \ \ systems of the hydrodynamic type and the Bogolubov-Whitham

\ \ \ \ \ \ averaging method, Dokl. Akad. Nauk SSSR, 270 (1983) No.4,
781-785.

[\textbf{3]}. S.P.Tsarev, Poisson brackets and one-dimensional Hamiltonian

\ \ \ \ \ \ systems of hydrodynamic type. Russian Akad.Sci.Dokl.Math.

\ \ \ \ \ 282 (1985),\ No.3, 534-537.

[\textbf{4]}. Mokhov O.I., Ferapontov E.V., Hamiltonian operators of

\qquad Hydrodynamic type associated with constant curvature metrics,

\qquad Usp.Math.Nauk, v.50, No.3 (1990), 191-192, translated as Russian

\ \ \quad \thinspace \thinspace\ Math.Surveys.

\textbf{[5]}. Maxim V.Pavlov, Elliptic coordinates and multi-Hamiltonian

\ \ \ \ \ \ structures of systems of hydrodynamic type, Russian Acad.Sci.

\ \ \ \ \ \ Dokl.Math. Vol.50 (1995), No.3, 374-377.

\textbf{[6]}. Ferapontov E.V., Nonlocal Hamiltonian Operators of Hydrodynamic

\qquad \ Type: Differential Geometry and Applications, Amer. Math. Soc.

\qquad\ Transl. (2) Vol.170, 1995.

\textbf{[7]}. Maxim V.Pavlov, Relationships between Differential
Substitutions and

\qquad Hamiltonian Structures of Korteweg-de Vries equation. Phys.Lett.A.

\qquad \textit{243 (1998) 295-300.}

\textbf{[8]}. O.I.Mokhov, Differential Geometry of Symplectic and Poisson

\ \ \ \ \ \ Structures on Loop Spaces of Smooth manifolds, and Integrable

\ \ \ \ \ \ Systems, Proceedings of the Steklov Institute of Mathematics,

\ \ \ \ \ V.217 (1997), pp.91-125.

\textbf{[9]}. A.B.Borisov, M.Pavlov, S.Zykov, The Kaup-Boussinesq system and

\ \ \ \ \ Prolifiration cheme'' to be published in Physica D, 2001.

\textbf{[10]}. Ferapontov E.V., Differential geometry of nonlocal Hamiltonian

\qquad \thinspace \thinspace \thinspace operators of hydrodynamic type,
Funktional. Anal. i Prilozhen.25

\qquad \thinspace \thinspace \thinspace (1991), No.3, 37-49; English transl.
in Functional Anal. Appl. 25

\qquad \ (1991).

\textbf{[11]}. Enriquez B., Orlov.A., Rubtsov V., Higher Hamiltonian
structures

\qquad \thinspace \thinspace on the KdV phase space (the sl(2) case).

\textbf{[12]}. Maxim V.Pavlov, Four Lagrangian representations for the
extended

\ \ \ \ \ \ \ \ Harry-Dym equation. to be published.

\end{document}